\documentclass[12pt]{iopart}

\usepackage{graphicx,amssymb}
\expandafter\let\csname equation*\endcsname\relax
\expandafter\let\csname endequation*\endcsname\relax

\usepackage[normalem]{ulem}
\usepackage[numbers,sort&compress,square]{natbib}
\usepackage{xcolor}
\usepackage{amsmath}
\usepackage{color}
\usepackage{xspace}
\usepackage{subfigure}
\usepackage{mathtools}
\usepackage{hyperref}
\usepackage{url}
\usepackage{orcidlink}
\usepackage{iopams}
\DeclarePairedDelimiter\abs{\lvert}{\rvert}%
\DeclarePairedDelimiter\norm{\lVert}{\rVert}%

\begin{document}

\title[SVD Based Discriminator for Binary Black Hole Searches]{Improved Binary Black Hole Search Discriminator from the Singular Value Decomposition of Non-Gaussian Noise Transients}

\newcommand{\IUCAA}{Inter-University Centre for Astronomy and Astrophysics, Post Bag 4, Ganeshkhind, Pune 411 007, India}
\newcommand{\WSU}{Department of Physics \& Astronomy, Washington State University, 1245 Webster, Pullman, WA 99164-2814, USA} 
\newcommand{\UWA}{Department of Physics, University of Western Australia, Crawley, WA 6009, Australia}

\newcommand{\ICRR}{Institute for Cosmic Ray Research, The University of Tokyo, 5-1-5 Kashiwanoha, Kashiwa, Chiba 277-8582, Japan}

\author{Tathagata Ghosh~\orcidlink{0000-0001-9848-9905}$^{1,2}$, Sukanta Bose~\orcidlink{0000-0002-4151-1347}$^{3}$, Sanjeev Dhurandhar$^2$, and Sunil Choudhary~\orcidlink{0000-0003-0949-7298}$^4$}

\address{$^1$~\ICRR}
\address{$^2$~\IUCAA}
\address{$^3$~\WSU}
\address{$^4$~\UWA}

\begin{abstract}

The sensitivity of current gravitational wave (GW) detectors to transient GW signals is severely affected by a variety of non-Gaussian and non-stationary noise transients, such as the blip, tomte, koi fish, and low-frequency blip `glitches'. These glitches 
share some time-frequency resemblance with GW signals from binary black holes. In earlier works~\cite{Joshi:2020eds, Choudhary:2022nvs}, the authors presented 
a method for constructing 
a $\chi^2$-distributed optimized statistic, based on the unified formalism of $\chi^2$ discriminators~\cite{Dhurandhar:2017aan}, to distinguish the blip glitches from the compact binary coalescence (CBC) signals. 
Unlike past works, the new $\chi^2$ discriminator is constructed from the most significant singular vectors obtained from the singular value decomposition of different classes of glitches in real detector data.
We find that the chi-square developed in this work performs as efficiently as in~\cite{Choudhary:2022nvs}, 
which used sine-Gaussian basis vectors.
This result supports past empirical findings that these glitches are reasonably well-modeled by sine-Gaussians. 
It also introduces a method for constructing signal- and glitch-based $\chi^2$ discriminators by directly using real data containing the glitches and, thus, holds promise for extensions to glitches that are captured less well by sine-Gaussians or other analytical functions.

\end{abstract}

\maketitle

\section{Introduction} \label{sec:introduction}

The Advanced Laser Interferometer Gravitational-wave Observatory (LIGO)~\cite{LIGOScientific:2014pky} and Advanced Virgo~\cite{VIRGO:2014yos} detectors have already observed more than $200$ gravitational wave (GW) sources up to the first part of the fourth observational run (O4)~\cite{LIGOScientific:2025slb}.
Most of the detections comprise binary black hole (BBH) events~\cite{KAGRA:2021vkt}. 
The search for BBHs in  GW detectors involves matched filtering their noisy strain data with templates modeled after theoretical BBH waveforms to construct a cross-correlation statistic that is optimal when that noise is Gaussian with a vanishing mean~\cite{Sathyaprakash:1991mt}.
However, non-Gaussian noise transients and the non-stationarity of detector noise~\cite{Abbott:2016xvh} can occasionally trigger templates with significant signal-to-noise ratios (SNRs)~\cite{LIGOScientific:2016gtq, Nuttall:2015dqa} and, consequently, produce false alarm triggers. These noise triggers reduce the sensitivity of BBH searches.
Predominantly, the high mass compact binary coalescence (CBC) searches are adversely affected by multiple types of noise transients~\cite{LIGOScientific:2017tza, Nitz:2017lco}. 

There are different types of glitches present in GW detectors. Some of these glitches arise from known instrumental and environmental sources, but others are of uncertain or unknown origins~\cite{Davis:2022dnd, 2018RSPTA.37670286N, Cabero:2019orq}. In this study, we focus on four common glitch classes: blip~\cite{Cabero:2019orq}, tomte, koi fish, and low-frequency blip~\cite{Bahaadini:2018git}.
A blip is a short-duration non-Gaussian artifact, lasting $\mathcal{O}(10)$ ms, with a moderately large frequency bandwidth, $\mathcal{O}(100)$ Hz, that frequently triggers BBH. 
This triggering is particularly accentuated in the search for high-mass CBC signals~\cite{Cabero:2019orq}, as the time-frequency evolution of the blip glitch closely resembles that of high-mass CBC signals. The blip's physical origin 
is poorly understood~\cite{Cabero:2019orq}. On average, the Advanced LIGO data contained approximately two blip glitches per hour of data during its first and second observing runs~\cite{Cabero:2019orq}. 
Tomte glitches~\cite{Alvarez-Lopez:2023dmv} typically occur at lower frequencies and have a somewhat longer duration than blip glitches. Koi fish glitches exhibit a bandwidth similar to that of blips but are characterized by additional spectral features on either side, including a prominent structure around 60 Hz that forms the distinctive ‘fins’ of the fish. Low-frequency blips have a similar rounded morphology to blips but do not extend to as high frequencies.
To distinguish short-duration glitches from CBC signals, chi-square tests, such as the power $\chi^{2}$~\cite{Allen:2004gu}, sine-Gaussian $\chi^{2}$~\cite{Nitz:2017lco}, and autocorrelation $\chi^{2}$~\cite{messick_gstlal, Chu:2020pjv}, are typically employed. 
In addition to the matched-filter SNR, the signal consistency test~\cite{Allen:2005fk} and the consistency of the phases, amplitudes, and times-of-coalescence among the near-concurrent GW signal candidates in the detector network, as expected of their common binary origin, are used in ranking them~\cite{Nitz:2017svb}.
In addition to $\chi^{2}$ discriminators, data-quality vetoes~\cite{LIGOScientific:2017tza} and gating~\cite{Usman:2015kfa} are employed to remove data segments contaminated by non-stationary behavior to improve the search sensitivity.
Machine–learning-based approaches have also been developed for glitch identification and classification (e.g., ~\cite{Zevin:2016qwy, Soni:2021cjy, Zevin:2023rmt, Wu:2024tpr}).

Apart from the power $\chi^{2}$, sine-Gaussian $\chi^{2}$, and autocorrelation $\chi^{2}$, a couple of improved discriminators 
were constructed~\cite{Joshi:2020eds, Choudhary:2022nvs} 
that exploited specifically how blips triggered BBH templates. Most recently, the optimized sine-Gaussian $\chi^{2}$ statistic~\cite{Choudhary:2022nvs}, based on the formalism of the unified $\chi^{2}$ discriminator~\cite{Dhurandhar:2017aan}, was shown to perform better than others for distinguishing spin-aligned BBH signals from blips. In~\cite{Joshi:2020eds, Choudhary:2022nvs}, the authors used the sine-Gaussian waveform as a model for the blip glitches to construct the new $\chi^{2}$. In contrast, this work \emph{does not use any preconceived glitch model} but employs the glitches themselves, obtained from the detector data, to develop their function space.

The function space associated with a particular class of glitches is constructed by applying singular value decomposition (SVD) to time snippets of real GW strain data containing them. The resulting singular vectors are then used to formulate their respective $\chi^{2}$ statistics.
In contrast to employing fiducial functions, such as sine-Gaussians, to capture the projections of glitches, the advantage of the new strategy is that the SVD vectors are specifically tailored to those glitches and, hence, can reduce the false dismissal of true CBC signals.
By supplementing real glitch data with simulated spin-aligned BBH injections added to real GW data, we demonstrate that this new SVD-based \textit{$\chi^{2}$} statistic performs better than other discriminators for BBH searches for a wide range of masses.
When the $\chi^{2}$ statistic is constructed using singular vectors derived from blip, tomte, koi fish, and low-frequency blip glitches, we refer to them as the \textit{blip} $\chi^{2}$, \textit{tomte} $\chi^{2}$, \textit{koi fish} $\chi^{2}$, and \textit{low-frequency blip} $\chi^{2}$, respectively. 
We also combine the singular vectors from the four glitch classes considered in this study to construct a common set of vectors, which can be applied across these classes. We refer to the resulting $\chi^{2}$ as the \textit{generic} $\chi^{2}$, since it is not tailored to any specific glitch class.
Any one of these $\chi^{2}$ discriminators that employs SVD vectors corresponding to glitches will be termed as an SVD-based $\chi^{2}$ statistic.
This approach has the added attraction that it may be useful in defining \textit{$\chi^{2}$} discriminators for glitches that cannot be resolved in terms of any known analytical basis functions.

The paper is structured as follows. In sections~\ref{match_filter_review} and~\ref{svd_review}, we briefly review the concepts of matched filtering and SVD, respectively. They are important tools used in this work. We also present a brief discussion of the fundamental concepts of the unified $\chi^{2}$ formalism in section~\ref{chisq_review}. 
Subsequently, in section~\ref{chisq_basis}, we detail the methodology developed for constructing the basis vectors, which is used to compute the glitch-specific $\chi^{2}$ statistics, and present the corresponding results in section~\ref{results}.
Finally, we summarize our findings and outline future research directions in section~\ref{conclusion}.

\section{Matched filtering} \label{match_filter_review}

The CBC detection statistic is based on the matched-filtering technique to identify the GW signals from the noise background \cite{Allen:2005fk, Dhurandhar:1992mw, Owen:1998dk}. 
The detector output $x(t)$ is the sum of GW signal $s(t)$ and the noise $n(t)$, which we take to be additive:
\begin{equation}
    x(t) = s(t) + n(t) \,.
\end{equation}
Assuming zero mean, stationary, and Gaussian noise, we have the usual relations:
\begin{equation} \label{eq:noise}
    \langle\widetilde{n}(f)\rangle=0, \ \ \ \  \langle\widetilde{n}^{\ast}(f) \widetilde{n}^{\prime}(f)\rangle=\frac{1}{2} S_{h}(f) \delta(f-f^{\prime}) \,,
\end{equation}
where the tilde over a quantity  denotes its Fourier transform, the angular brackets denote the ensemble average, 
$S_{h}(f)$ is taken to be the time-independent, single-sided noise power spectrum density (PSD), and $\delta$ is the Dirac delta function.
Here, we follow the convention that for any function in the time domain,  $q(t)$, the Fourier transform is given by $\widetilde{q}(f)=\int_{-\infty}^{\infty} dt ~e^{2\pi ift} q(t)$. 
Now, before going into the details of the matched-filtering, the (weighted) inner product of two functions, $x(t)$ and $y(t)$, can be introduced as follows~\cite{Finn:1992wt}:
\begin{equation} \label{eq:inner_prod}
    (\mathbf{x} , \mathbf{y}) = 4 \Re \int_{0}^{\infty} \frac{\widetilde{x}^{\ast}(f)\widetilde{y}(f)}{S_{h}(f)} df \\.   
\end{equation}
In the context of GW searches, the data $x(t), y(t), ...$,
acquired as discrete time series, over a time interval $[0, T]$, can be considered as vectors in the space $\mathcal{D}$ of data time series, and denoted by the boldface letters $\mathbf{x}$, $\mathbf{y}$, etc. With the scalar product defined in equation~(\ref{eq:inner_prod}), 
$\mathcal{D}$ is just the Hilbert space of data trains $\mathbf{x}$ over the time interval $[0, T]$. 
Note that we often work with whitened data, defined as $\widetilde{x}_{w}(f) \equiv \widetilde{x}(f)/\sqrt{S_{h}(f)}$. In this case, the inner product in equation~\ref{eq:inner_prod} takes the form of a Euclidean scalar product of the whitened quantities.

Matched filtering is performed by taking the inner product (equation~\eqref{eq:inner_prod}) of the GW signal $\mathbf{x}$ with each of the normalized templates $\mathbf{h}$ to calculate the matched-filter SNR. Since the initial phase of the GW signal in the strain data is unknown, the matched-filter SNR is computed as a function of time by maximizing over the phase of the signal. The CBC search identifies the times when the matched-filter SNR exceeds a predetermined threshold. Since the parameters of the detected GW signal are unknown {\it a priori}, a large set of waveform templates that densely cover the parameter space is employed.

\section{The unified $\chi^{2}$ formalism} \label{chisq_review}

The mathematical framework for the unified $\chi^{2}$ discriminator in the context of CBC searches is detailed in~\cite{Dhurandhar:2017aan}. 
The fundamental criteria of the $\chi^{2}$ statistic is that it equals zero for the CBC signal $\mathbf{s}
\equiv A\mathbf{h}$ (i.e., the amplitude $A$ multiplied by the normalized waveform $\mathbf{h}$) and follows the $\chi^{2}$ distribution for Gaussian noise. However, the detector data includes the CBC signal and the noise $\mathbf{n}$, implying that for additive noise the strain signal is $\mathbf{x} = \mathbf{s} + \mathbf{n}$. 

In this framework, the basic idea is that any $\chi^{2}$ can be associated with a set of vectors orthogonal to the signal, which span a subspace $\mathcal{S}^{\perp}$ of $\mathcal{D}$. In practice, in order to keep the computational cost under control, we desire that the subspace $\mathcal{S}^{\perp}$ has a small number of dimensions---these will constitute the degrees of freedom (DOFs) of the $\chi^2$. We may express $\mathcal{D}$ as a direct sum of $\mathcal{S}$ and $\mathcal{S}^{\perp}$, i.e., $\mathcal{D} = \mathcal{S} \oplus \mathcal{S}^{\perp}$. The data vector $\mathbf{x} \in \mathcal{D}$ can be decomposed into two parts as $\mathbf{x} = \mathbf{x}_{\mathcal{S}} + \mathbf{x}_{\mathcal{S}^{\perp}}$,
where $\mathbf{x}_{\mathcal{S}}$ and $\mathbf{x}_{\mathcal{S}^{\perp}}$ represent the projections of $\mathbf{x}$ into subspaces $\mathcal{S}$ and $\mathcal{S}^{\perp}$ respectively. 
Then, the $\chi^{2}$ statistic is defined as:
\begin{equation} \label{chisq_def}
    \chi^{2}(\mathbf{x}) = \norm{\mathbf{x}_{\mathcal{S}^{\perp}}}^{2} \,.
\end{equation}
To proceed, we choose a collection of orthonormal basis vectors $\mathbf{e}_{\alpha}$ in $\mathcal{S}^{\perp}$ with dimension $p$ (and $\alpha=1,2,...,p$), such that $(\mathbf{e}_{\alpha}, \mathbf{e}_{\beta}) = \delta_{\alpha \beta}$, which is just the Kronecker delta. The $\chi^{2}$ statistic has following properties:
\begin{enumerate}
    \item For any data vector $\mathbf{x} \in \mathcal{D}$, we can write from equation~\eqref{chisq_def}

    \begin{equation} \label{chisq_generic}
        \chi^{2}(\mathbf{x}) = \sum_{\alpha=1}^{p} \abs{(\mathbf{x}, \mathbf{e}_{\alpha})}^{2}\,.
    \end{equation}
    \item Since $\mathbf{h}$ has no projection onto $\mathcal{S}^{\perp}$, i.e., $\mathbf{h}_{\mathcal{S}^{\perp}}=0$, it is evident that $\chi^{2}(\mathbf{h})=0$. 
    \item If the noise $\mathbf{n}$ is assumed to be Gaussian and satisfies equation~\eqref{eq:noise}, then 
    \begin{equation}
        \chi^{2} (\mathbf{n}) = \norm{\mathbf{n}_{\mathcal{S}}}^{2} = \sum_{\alpha=1}^{p} \abs{(\mathbf{n}, \mathbf{e}_{\alpha})}^{2}
    \end{equation}
    follows $\chi^{2}$ distribution with $p$ DOFs.
\end{enumerate}
One may choose to work with any orthonormal basis of $\mathcal{S}^{\perp}$ in this formalism. In such a basis, the statistic $\chi^{2}$ manifestly follows a $\chi^2$ distribution since it is expressed as a sum of squares of independent Gaussian random variables with zero mean and unit variance.

In searching for a CBC signal in noisy data, we must consider a template bank consisting of a family of signal waveforms that depend on several parameters, such as masses, spins, and other kinematical parameters. The templates in the bank are normalized and densely cover a manifold $\mathcal{P}$ of waveforms, which is a sub-manifold of $\mathcal{D}$.
Each point of $\mathcal{P}$ represents a normalized waveform, and $\mathcal{S}^{\perp}$ can be associated with each point of $\mathcal{P}$. Therefore, a general $\chi^2$ has the structure of a vector bundle, with $\mathcal{P}$ as the base space and $\mathcal{S}^{\perp}$ as the fibre~\cite{dhurandhar_book}.
One can thus use this general mathematical framework to construct a $\chi^{2}$ discriminator that best distinguishes the signals and glitches in any given observation data set.

\section{Singular value decomposition (SVD)} \label{svd_review}

In a typical observation run, one encounters BBH template triggers from thousands of glitches, which are vectors in $\mathcal{D}$. In contrast, the number of DOFs (or, equivalently, basis vectors) of the $\chi^2$ is typically much smaller in order to contain computational cost. The glitches, therefore, span a high-dimensional space; it helps to whittle this down to a lower-dimensional space that continues to approximate the space of those glitches well. Basically, the glitches must project substantially on this smaller dimensional space.
This projection helps reduce the cost of computing the $\chi^2$. The goal is achieved by applying the SVD procedure to a subset of glitches from each glitch class, thereby constructing a lower-dimensional space for that class. This section briefly reviews the SVD algorithm, an important tool used in this study. The SVD algorithm factors a matrix $B$ into a product of three matrices, 
\begin{equation} \label{matrix_svd}
    \mathbf{B} = \mathbf{U} \mathbf{\Sigma} \mathbf{V}^{\dag} \,,
\end{equation}
or, in terms of elements of matrices,
\begin{equation} \label{matrix_svd_element}
    b_{\mu j}=\sum_{\nu} u_{\mu\nu}\sigma_{\nu}v^*_{\nu j} \\.
\end{equation}
In equation~\eqref{matrix_svd}, $\mathbf{V}$ is a matrix of right singular vectors that form an orthonormal basis set; $\mathbf{\Sigma}$ is a diagonal matrix with positive real numbers $\{\sigma_{\nu}\}$, also known as singular values, arranged in descending order of magnitude; the rows of $\mathbf{V}^{\dag}$ (the  dagger $\dag$ in the superscript denotes the Hermitian conjugate) are the singular vectors, $\vec{v}_{\mu}$, satisfying,
\begin{equation}
    \sum_{j} v_{\mu j}^* v_{\nu j} = \delta_{\mu \nu} \\.
\end{equation}

The column vectors of $\mathbf{V}$ are termed as singular vectors, and they form an orthonormal basis of $\mathbf{V}$. The Eckart-Young-Mirsky theorem~\cite{Eckart1936TheAO} states that the best $k$-dimensional approximation to the vector space spanned by the row vectors of $\mathbf{B}$ is obtained by taking the first $p$ singular vectors. The value of $p$ is chosen to obtain the desired accuracy, which can be decided from the Frobenius norm. It is calculated from the singular values as follows:
 \begin{equation} \label{eq:accuracy}
     \norm{\mathbf{B}}_{F}^{2} = \sum_{i=1}^{M} \sum_{j=1}^{N} \abs{b_{ij}}^{2} = \sum_{k=1}^{r} \sigma_{k}^{2} \,.
 \end{equation}
For instance, if we require $\eta~\%$ level of accuracy, $p$ should be chosen so that \break $\sum_{k=1}^{p} \sigma_{k}^{2} \gtrsim 0.01 ~\eta~ \norm{\mathcal{B}}_{F}^{2}$.
We apply the SVD algorithm to a pre-selected set of glitches from each class, drawn from {\it real detector data,} to investigate their underlying time-frequency morphology. 

However, there is one more consideration: the usual SVD algorithm assumes the Euclidean scalar product. We see that when working with the real GW strain data, including CBC signals and different classes of glitches, the scalar product includes the inverse of PSD as a weight. So, we perform SVD on the whitened strain data.

\section{Constructing SVD-based $\chi^{2}$ discriminators} \label{chisq_basis}

In the preceding section, we briefly reviewed the general mathematical framework for constructing any $\chi^{2}$ discriminator in  CBC signal searches. This section applies that methodology to construct a  $\chi^{2}$ to specifically discriminate blip, tomte, koi fish, and low-frequency blip glitches from BBH signals effectively. It involves constructing a set of basis vectors that span a low-dimensional subspace of $\mathcal{D}$, based on the framework of the generic $\chi^{2}$ discriminator~\cite{Dhurandhar:2017aan}, as discussed in section~\ref{chisq_review}. A few earlier approaches~\cite{Joshi:2020eds, Choudhary:2022nvs} based on the unified $\chi^{2}$ formalism efficiently employ the sine-Gaussian model of blips to construct the $\chi^2$. In contrast, this work utilizes {\it the blip, tomte, koi fish and low-frequency blip} glitches in real detector data to construct the $\chi^2$. The key idea involves the construction of a function space based on the time-domain characteristics of the glitches.
To avoid over-fitting the $\chi^2$ statistic to the glitches, we used a relatively few (i.e., $\sim 100$) glitches from each class to construct the corresponding function space.
Subsequently, this function space is used to compute the new $\chi^{2}$ on larger sets of glitches, comprising $\sim 3000$ blips, $\sim 700$ tomtes, $\sim 1200$ koi fishes, and $\sim 1300$ low-frequency blips.
The assumption is that a reasonably sized sample of glitches is a good representation of all glitches of that kind.
Essentially, the glitches cover a relatively small subspace of the data vector space.

\begin{figure*}
    \centering
    \includegraphics[scale=0.5]{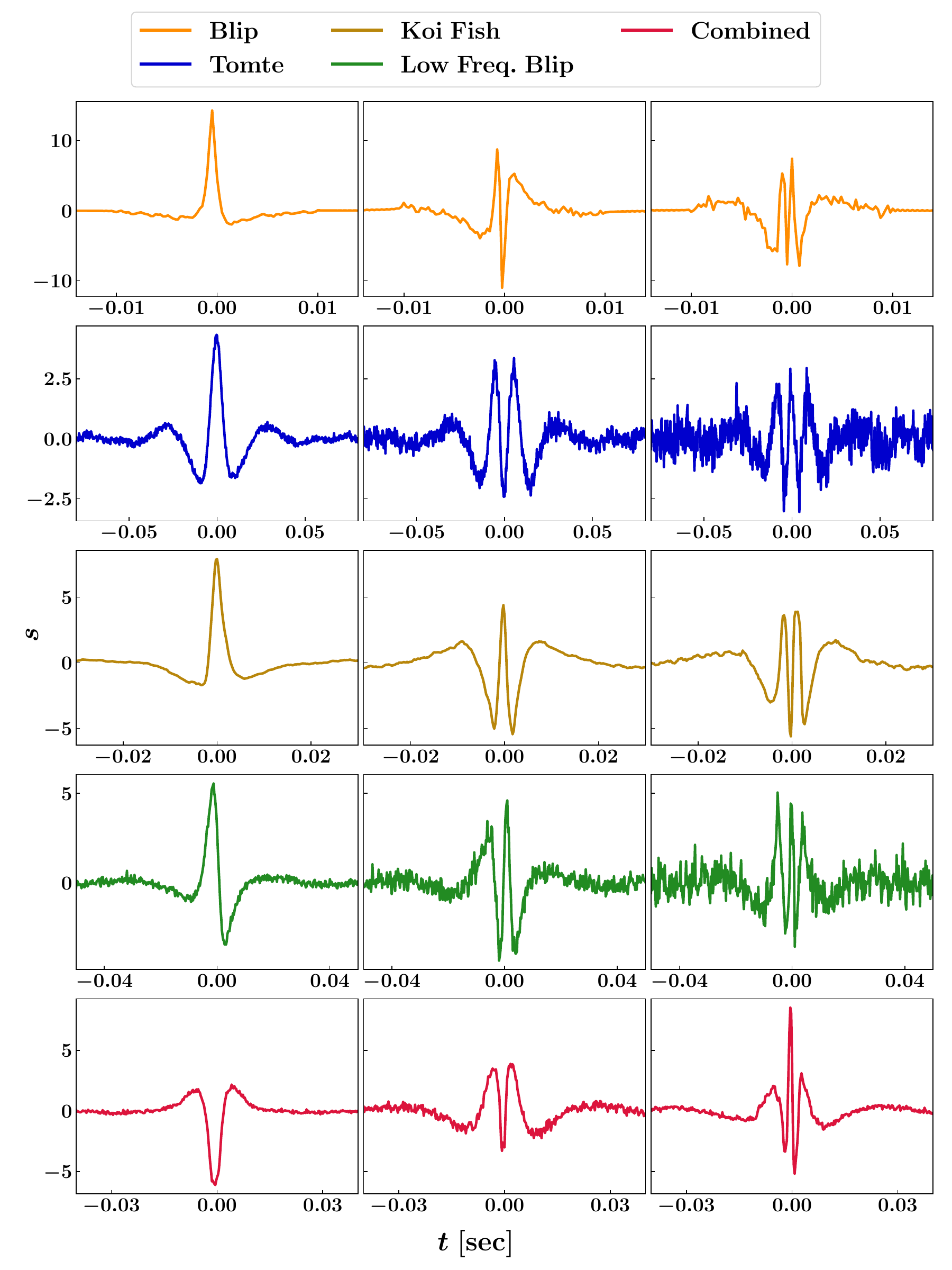}
    \caption{The first four rows show the first three singular vectors obtained from $100$ glitches of each class---blip, tomte, koi fish, and low-frequency blip, respectively. The bottom row shows the first three singular vectors obtained by performing an SVD on the $12$ singular vectors displayed in the panels above.}
    \label{fig:sv_H1}
\end{figure*}

This study considers the glitches from the LIGO Hanford (H1) detector data during the first part of the third observation (O3a) run~\cite{glanzer_jane_2021_5649212}, accessed through the Gravitational Wave Open Science Center~\cite{KAGRA:2023pio}. These glitches are identified by Gravity Spy~\footnote{publicly available at \href{https://zenodo.org/records/5649212}{Zenodo}~\cite{glanzer_jane_2021_5649212}}~\cite{Glanzer:2022avx} with a confidence level of $90\%$ or higher. We use $16$ sec strain data sampled at $4096$ Hz. The data has been downloaded using \verb+GWpy+~\footnote{\url{https://gwpy.github.io/docs/ma chi squarelatest/}}~\cite{gwpy} with the glitch positioned at the center of the data snippet. Although the durations of glitches are on the order of a few milliseconds to tens of milliseconds, we consider $16$ seconds of data to ensure enough sample points in the frequency domain data. 

We select $100$ glitches from each class, from the early part of H1 data during the O3a run for performing the SVD, and find that the top three singular values are much larger than the rest. Those singular values correspond to three singular vectors that are found to capture over 80\% of the power of these glitches, and effectively describe the glitches' time-frequency power distribution. Since our SVD vectors are constructed from {\em real} glitches, we avoid data snippets with known additional noise artifacts in them, lest they contaminate the basis vectors so deduced.
We, therefore, zero-pad the whitened strain data,  retaining an appropriate time snippet around the central time of each glitch, depending on its class.
We took $16$ s of data in the neighborhood of the glitch, excluding the glitch itself, and applied Welch’s method with median averaging~\cite{1161901} to compute the PSD. The resulting PSD was then used in analyzing the data with the glitch.

The frequency-domain strain data thus prepared are next used to construct the matrix $\mathcal{B}$. In this so-called {\em glitch}
matrix, each row corresponds to a particular glitch, and each column corresponds to a specific frequency index. Subsequently, we perform the SVD on this matrix.
In figure~\ref{fig:sv_H1}, the top four rows show the first three singular vectors of the blip, tomte, koi fish, and low-frequency blip glitches, respectively.~\footnote{The SVD is performed in the Fourier domain, but the singular vectors are plotted in the time domain for visualization purposes.}
We employ only the first three singular vectors for each glitch type, which capture $\gtrsim 80\%$ of the corresponding glitch features (see equation~\eqref{eq:accuracy}), ensuring sufficient representation of their morphologies.
As detailed below, we construct the glitch-specific $\chi^{2}$ by using the three leading singular vectors derived from each class---blip, tomte, koi fish, and low-frequency blip. 
Additionally, we perform SVD on the combined set of those $12$ singular vectors
and select the $3$ singular vectors output by that process corresponding to the top three singular values. 
These three singular vectors capture about $68\%$ of the morphology across all glitch classes considered in this study.
This corresponding set is shown in the last row of figure~\ref{fig:sv_H1}, and will be used below to construct the generic $\chi^2$ to discriminate blip, tomte, koi fish, and low-frequency blip glitches from CBC signals.
This approach is particularly valuable because it is agnostic to the glitch type, unlike glitch-specific $\chi^{2}$ statistics that are constructed using class-dependent information.

Since the $\chi^{2}$ construction requires that the  CBC template triggered by the GW data is orthogonal to the three SVD basis vectors, we construct {\em truncated} basis vectors by subtracting from each singular vector its component along the triggered CBC template.
Due to the morphological distinctions between the CBC signal and the glitch, the subtraction of the triggered template corresponding to the glitch is expected to have a negligible impact on the $\chi^{2}$ value. Nevertheless, the CBC template has significantly diminished projection onto the truncated basis vectors.
We discuss the details of the template bank for performing matched-filtering and how the trigger time is chosen to calculate the glitch-specific, or generic $\chi^{2}$s in section~\ref{sec:chisq_prepare}.

In general, the truncated basis vectors thus produced would not remain orthogonal even though they were produced from ones that were.
Therefore, we apply the Gram-Schmidt orthogonalization procedure to the truncated basis vectors so that we have an orthonormal basis set. 
These vectors now serve as the final basis vectors for computing the proposed SVD-based $\chi^{2}$, on to which the CBC signals and the glitches are projected.

\section{Preparation for $\chi^{2}$ calculation} 
\label{sec:chisq_prepare}

To study how the discriminatory power of our $\chi^2$ statistic varies with the CBC masses, we use several aligned-spin template banks, which were constructed by employing the stochastic placement algorithm. These banks cover component masses ranging from $20~M_{\odot}$ to $70~M_{\odot}$, with each bank having a component-mass width of  $10~M_{\odot}$. The mass ranges are as follows: $20-30~M_{\odot}$, $30-40~M_{\odot}$, $...$, $60-70~M_{\odot}$. The aligned dimensionless black hole spin parameter in these banks varies between $0.0$ and $0.9$.
The template banks are constructed using the representative noise PSD of the H1 noise strain during the O3a run, taken on September 5, 2019~\cite{KAGRA:2013rdx}, and by setting the minimum match required of the nearest neighboring template to $97\%$. 

\begin{figure*}
    \centering
    \includegraphics[scale=0.38]{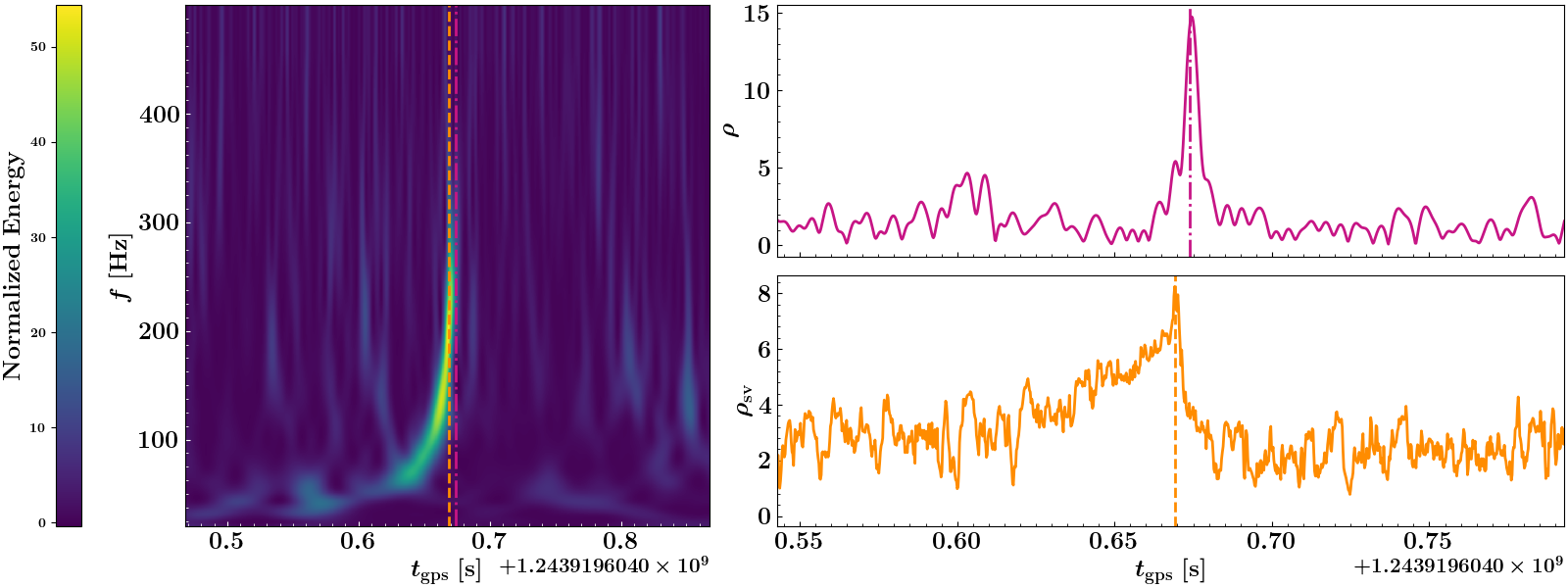}
     \caption{A simulated CBC signal added to real detector strain data. \emph{Top-right panel}: The SNR time series obtained by matched filtering real GW data with a simulated CBC signal added to it. The template used for producing it was the loudest one, in this time-stretch, from the aligned-spin template bank with component mass between $30~M_{\odot}$ and $40~M_{\odot}$. Here, the peak matched-filter SNR is $14.7$. \emph{Bottom-right panel}: SV-SNR (discussed in section~\ref{sec:chisq_prepare} time series for the same simulated CBC signal, with the peak SNR of $8.2$. Note that the time of the peak SNR is different in the two right plots (see~\cite{Bose:2016jeo} for an explanation).
    \emph{Left panel}: The time-frequency plot of a CBC signal is shown to illustrate the lag between the time the CBC template is triggered (indicated by the vertical dot-dashed magenta line, at a later time) and the blip singular vectors are triggered (shown by the vertical dashed orange line, at an earlier time).
    }
    \label{fig:cbc_trigger_time}
\end{figure*}

\begin{figure*}
    \centering
    \includegraphics[scale=0.38]{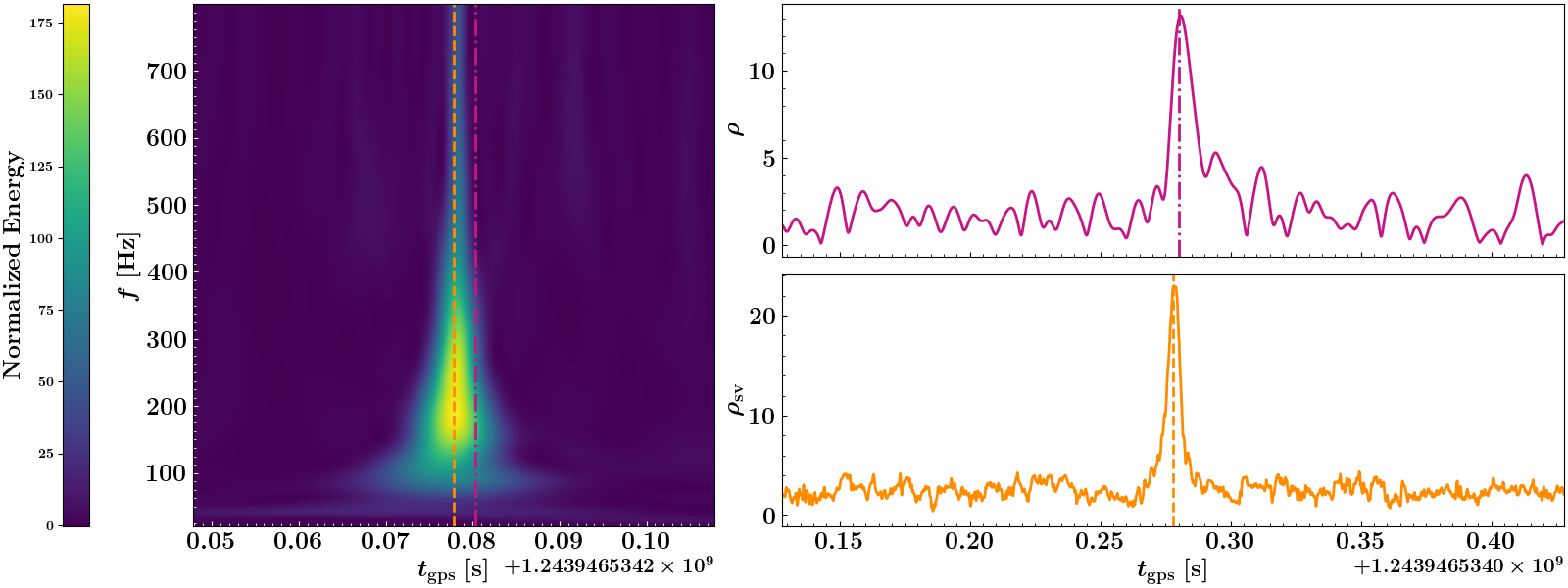}
    \caption{A blip glitch in real GW strain data. \emph{Left panel}: Time-frequency plot of the blip glitch.\emph{Top-right panel}: The SNR time series plot of the blip glitch obtained by matched-filtering with the loudest CBC template that was triggered by this time-stretch. The CBC template bank employed here was identical to the one used in figure~\ref{fig:cbc_trigger_time}. The peak value of the matched-filter SNR is $13.1$. \emph{Bottom-right panel}: 
    SV-SNR time series for the same blip glitch, with the peak SNR of $23$.}
    \label{fig:blip_trigger_time}
\end{figure*}

The CBC template banks are used to match filter and compute the SNR for the different strain data segments  that either
contain
a glitch (e.g., as noted by Gravity Spy) or have a simulated CBC signal injected.
On a given strain snippet, the CBC-template trigger time, also simply termed as the `trigger time', is identified as the time at which the SNR time-series attains maximum value---maximized over a bank of CBC templates. 
In contrast, we define the {\it singular-vector} trigger time similarly, except that instead of a CBC template, the three SVD vectors introduced in section~\ref{chisq_basis} are used to compute the `SV-SNR'. The SV-SNR is just the square root of the expression in equation~\eqref{chisq_generic}, with the $\mathbf{e}_{\alpha}$ being the three SVD vectors.
Figure~\ref{fig:cbc_trigger_time} (figure~\ref{fig:blip_trigger_time}) shows this SV-SNR time series obtained from matched filtering real GW strain data containing a BBH signal (blip glitch), both with the loudest CBC template it triggered as well as with the three singular vectors.

If a CBC template has the same intrinsic parameters as a loud CBC signal, then the trigger time is a good estimator of the signal end-time. Instead, if the strain data contains a sine-Gaussian `glitch', then the trigger time can be significantly different from its central time~\cite{Bose:2016jeo}.
The value of the SVD-based $\chi^2$ statistic is largest at that central time, which is better estimated by the singular-vector trigger time.
Figure~\ref{fig:blip_trigger_time} shows that the peak in the SNR time series obtained from matched filtering with the singular vectors identifies the blip time quite accurately compared to that obtained from matched filtering with the loudest CBC template. 
As shown there, the peak in the former time series lags behind the peak in the latter. We simply term this as the `glitch' time-lag, which is typically of the order of a few milliseconds to tens of milliseconds, depending on the glitch class.
In contrast, as shown in figure~\ref{fig:cbc_trigger_time}, the accuracy in determining the signal end time of a CBC signal may be slightly compromised when using singular vectors. 
Nevertheless, we persist with this procedure since we may not know in advance whether the data corresponds to a CBC signal or a glitch.

When computed on data with a true glitch, the SVD-based $\chi^{2}$ decreases rapidly with time. The value of this time-lag is used for determining the time-point at which the SVD-based $\chi^{2}$ is computed, which is consistent with the procedure that was employed in~\cite{Choudhary:2022nvs} for computing the optimized sine-Gaussian $\chi^{2}$.~\footnote{In~\cite{Choudhary:2022nvs}, the parameter analogous to the time-lag was called $t_d$---for the sine-Gaussian basis vectors employed there.}
Since {\it a priori} it would not be known if the trigger was caused by a CBC signal or a glitch, the SVD-based $\chi^{2}$ is always computed at the singular-vector trigger time. 
We use the same procedure to construct the generic $\chi^{2}$ for both CBC signals and glitches.
In contrast, note that the power and the sine-Gaussian $\chi^{2}$ are computed at the CBC template trigger times, both for CBC signals and glitches.

\section{Results} \label{results}

To examine the effectiveness of both the glitch-specific and generic $\chi^{2}$ discriminators, we simulated spin-aligned BBH signals using the \verb+IMRPhenomD+~\cite{Khan:2015jqa} waveform model with component masses binned into $5$ distinct sets, between $20~M_{\odot}$ and $70~M_{\odot}$, in steps of $10~M_{\odot}$.
Our analysis involved glitches from the four classes considered in this study---approximately $3000$ blips, $700$ tomtes, $1200$ koi fishes, and $1300$ low-frequency blips--recorded in the LIGO-Hanford detector during the O3a run. We injected the simulated CBC signals in real data near the glitches used in this work.

The signals simulated are consistent with CBCs distributed uniformly in source-frame time and comoving volume between $1$~Gpc and $3$~Gpc, with the constraint that the 
detector SNR
should be at least $4$.
We employ the $3$ most significant singular vectors that are constructed by performing SVD of $100$ glitches from the LIGO-Hanford O3a dataset, as described in section~\ref{chisq_basis}. 
We then followed the procedures described in sections~\ref{chisq_basis} and~\ref{sec:chisq_prepare} to calculate the glitch-specific $\chi^{2}$ as well as generic $\chi^{2}$ statistics for all glitches in each class and the injected CBC signals.
Since we use both the real and imaginary parts of the first $3$ singular vectors, the DOFs of the corresponding $\chi^{2}$ are $6$.
Additionally, we compute the power $\chi^{2}$, sine-Gaussian $\chi^{2}$, and optimized sine-Gaussian $\chi^{2}$ for the same glitches---blips and tomtes---and CBC signals. We consider $d=16$ bins in the frequency domain to compute the power $\chi^{2}$ between $20$ Hz and $2000$ Hz. So, the DOFs of the power $\chi^{2}$ test are $2d-2 = 30$. On the other hand, the DOF of the sine-Gaussian $\chi^{2}$ is $14$, arising from placing $7$ equispaced sine-Gaussian tiles in steps 
of $15$ Hz, and $15-120$ Hz above the final frequency of the triggered waveform template. 
We follow the methodology described in~\cite{Choudhary:2022nvs} to compute the optimized sine-Gaussian $\chi^{2}$ with DOFs equal to $6$.
In contrast to~\cite{Choudhary:2022nvs}, here we took the central frequency and the quality factor ranges in the construction of the optimized sine-Gaussian $\chi^2$
to be somewhat broader, namely, 
$f_{0} \in [20, 500]$ Hz and 
$Q \in [2, 12]$, respectively. 
These broader ranges are found to more completely capture the power across the different glitch classes considered in this study.

\begin{figure*}
    \centering
    \includegraphics[scale=0.41]{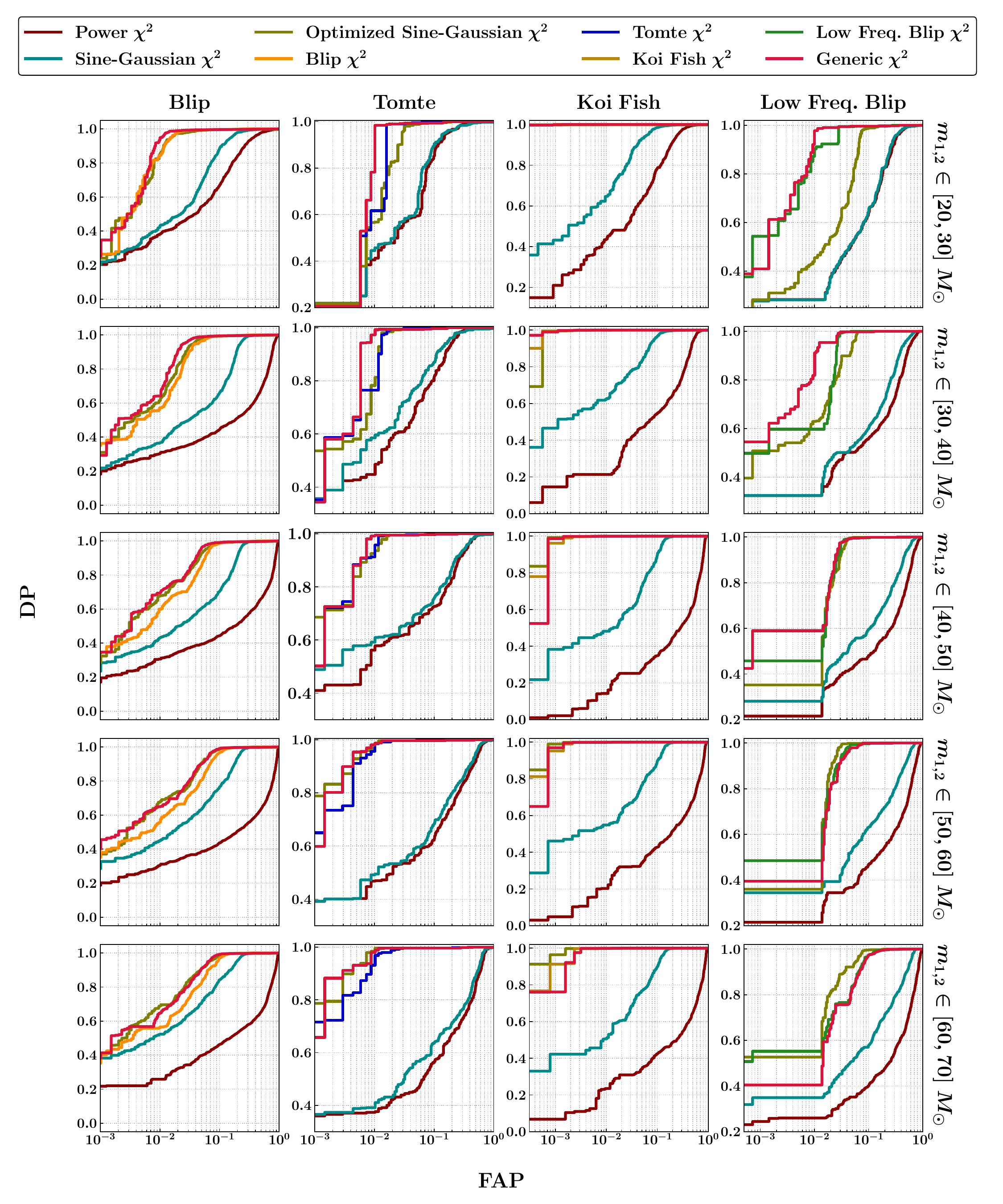}
    \caption{ROC curves showing the performance of various $\chi^{2}$ statistics for different glitch classes---blip, tomte, koi fish, and low-frequency blip (left to right columns)---evaluated using CBC signals in different mass ranges, as indicated on the right of the respective panels.}
    \label{fig:roc_compare}
\end{figure*}

\begin{figure*}
    \centering
    \includegraphics[scale=0.46]{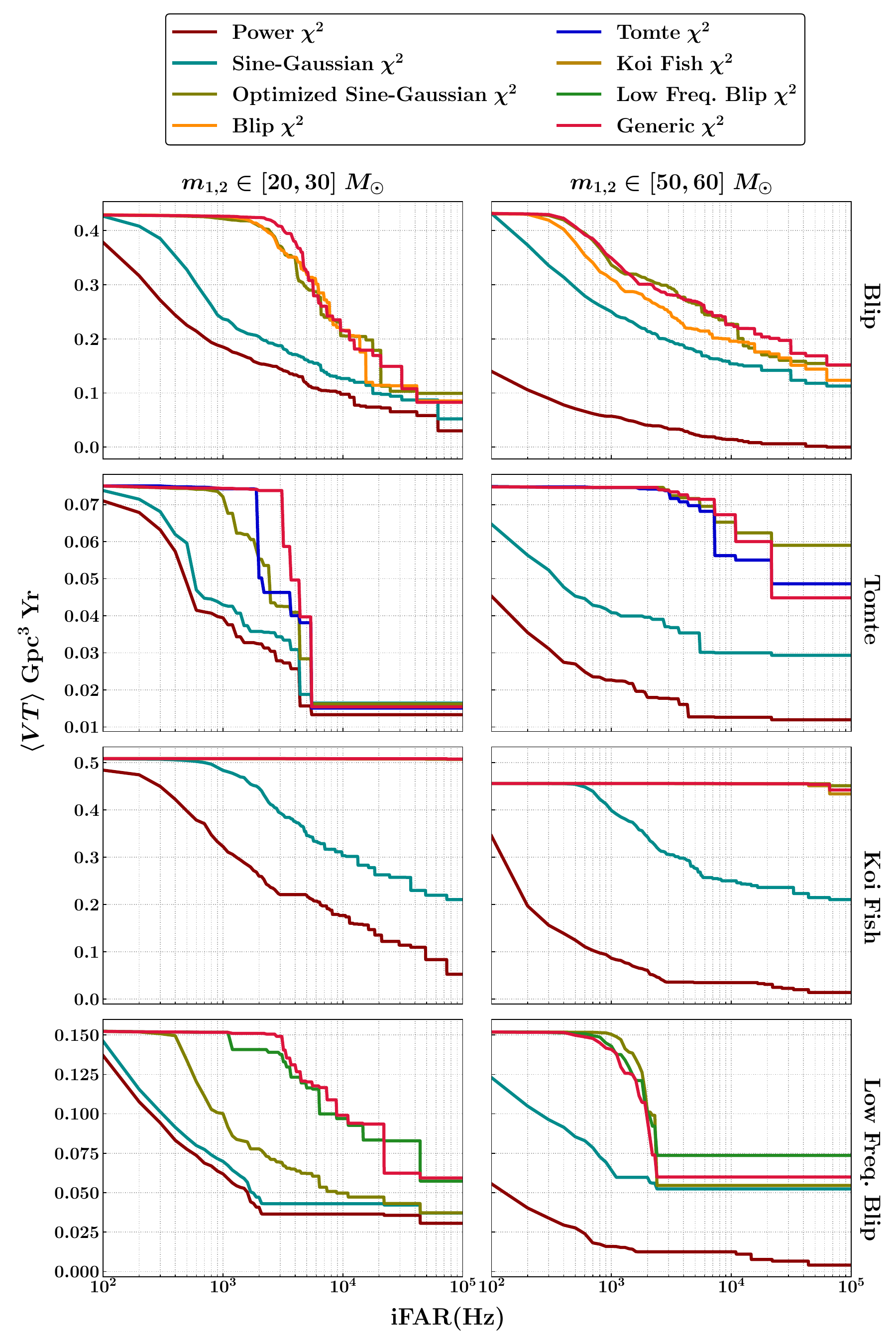}
    \caption{Volume–time sensitivity (VT) as a function of inverse false-alarm rate (iFAR) for different $\chi^{2}$ statistics, shown for two mass bins: $m_{1,2} \in [20,30]~M_{\odot}$ and  $m_{1,2} \in [50,60]~M_{\odot}$.}
    \label{fig:vt_compare}
\end{figure*}

\begin{figure*}
    \centering
    \includegraphics[scale=0.8]{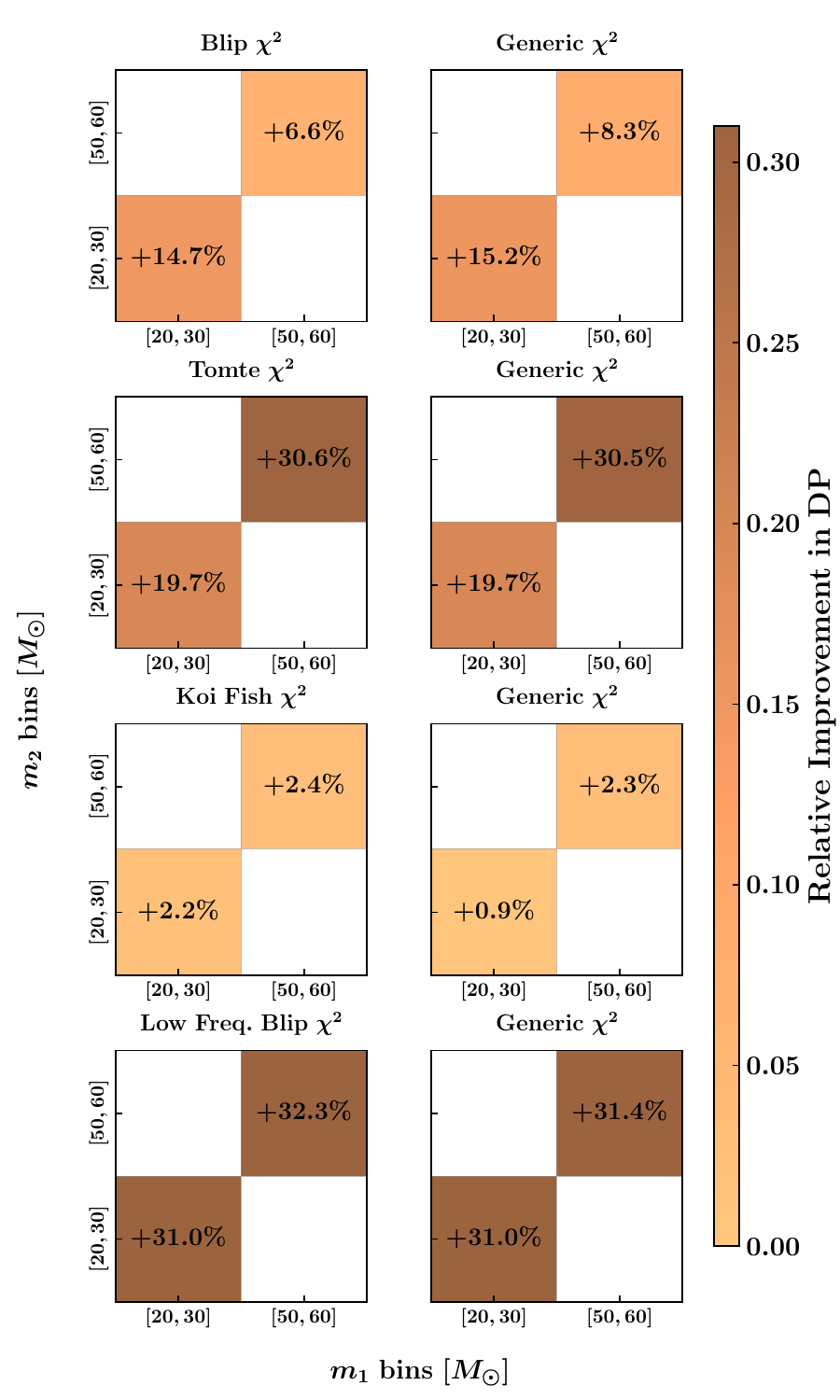}
    \caption{\emph{Left panel:} Improvement in detection probability (DP) for various glitch classes (blip, tomte, koi fish, and low-frequency blip), computed using a $\chi^{2}$ statistic built from glitch-specific singular vectors and compared to the sine-Gaussian $\chi^{2}$, using simulated CBC signals and real glitches.
    \emph{Right panel:} Same as the left panel, but using the $\chi^{2}$ constructed from common singular vectors, formed by combining the singular vectors from different glitch classes and applied across all glitches.}
    \label{fig:dp_plot}
\end{figure*}

In order to rank the triggers (glitches and CBC signals) based on power $\chi^{2}$, we employ the following re-weighted SNR statistics~\cite{LIGOScientific:2011jth},
\begin{equation} \label{snr_pow}
    \rho_{\rm pow} = 
\begin{cases}
    \rho ,&  \chi_{r}^{2} \leq 1\\
    \rho \left[\frac{1}{2}\left(1+\left(\chi_{r, {\rm pow}}^{2}\right)^{3}\right)\right]^{-1/6},  & \chi_{r}^{2} \geq 1 \,.
\end{cases}
\end{equation}
On the other hand, when employing the sine-Gaussian $\chi^{2}$, we use the following ranking statistic to rank the glitch and CBC triggers:
\begin{equation} \label{snr_sg}
    \rho_{\rm sg} \equiv 
\begin{cases}
    \rho_{\rm pow} ~,&  \chi_{r, {\rm sg}}^{2} \leq 4\\
    \rho_{\rm pow} \left(\chi_{r,{\rm sg}}^{2}/4 \right)^{-1/2} ~,  & \chi_{r, {\rm sg}}^{2} > 4 \,,
\end{cases}
\end{equation}
as was defined in~\cite{Nitz:2017lco}. 
We employ the same ranking statistic (equation~\eqref{snr_sg}) for the optimized sine-Gaussian $\chi^{2}$ as well as blip and tomte $\chi^{2}$s to rank their triggers~\cite{Choudhary:2022nvs}. 
Based on the $\chi^{2}$ and the corresponding re-weighted SNR, we construct the ROC curve as shown in figure~\ref{fig:roc_compare} to compare the efficacy of different $\chi^{2}$ in the identification of the glitches.
It is evident from figure~\ref{fig:roc_compare} that the glitch-specific $\chi^{2}$ and generic $\chi^{2}$ perform similar to the optimized sine-Gaussian $\chi^{2}$, but better than both the power $\chi^{2}$ and the sine-Gaussian $\chi^{2}$ in distinguishing identifying the glitches  against CBC signals.
We also compare the volume–time sensitivities~\cite{LIGOScientific:2016kwr, Tiwari:2017ndi} as functions of the inverse false-alarm rate for all the $\chi^{2}$ statistics considered in this study, as shown in figure~\ref{fig:vt_compare}, focusing on two representative mass bins.
The volume–time sensitivity is computed as the ratio of the number of recovered injections to the total number of injected signals, as outlined in~\cite{Tiwari:2017ndi}.
Additionally, figure~\ref{fig:dp_plot} shows the improvement in the true positive rate achieved by the proposed glitch-specific $\chi^{2}$ over the sine-Gaussian $\chi^{2}$ when the data contain glitches from the classes considered in this study or simulated CBC signals. 
A similar improvement is seen for the generic $\chi^{2}$ over the sine-Gaussian $\chi^{2}$ across data sets containing multiple glitch classes, in addition to simulated CBC signals.
For blip glitches, all $\chi^{2}$ statistics---optimized sine-Gaussian, blip, generic, and sine-Gaussian---perform comparably. This is likely because blip glitches resemble sine-Gaussian waveforms and typically occur at higher frequencies, where the sine-Gaussian $\chi^{2}$ is effective. 
In contrast, the optimized sine-Gaussian $\chi^{2}$, glitch-specific $\chi^{2}$,  and generic $\chi^{2}$ show significant improvement over the sine-Gaussian $\chi^{2}$ in identifying other glitches.
These glitches typically occur at lower frequencies and, therefore, are better captured by statistics that employ SVD to more accurately model those morphologies.

\begin{figure*}
    \centering
    \includegraphics[scale=0.7]{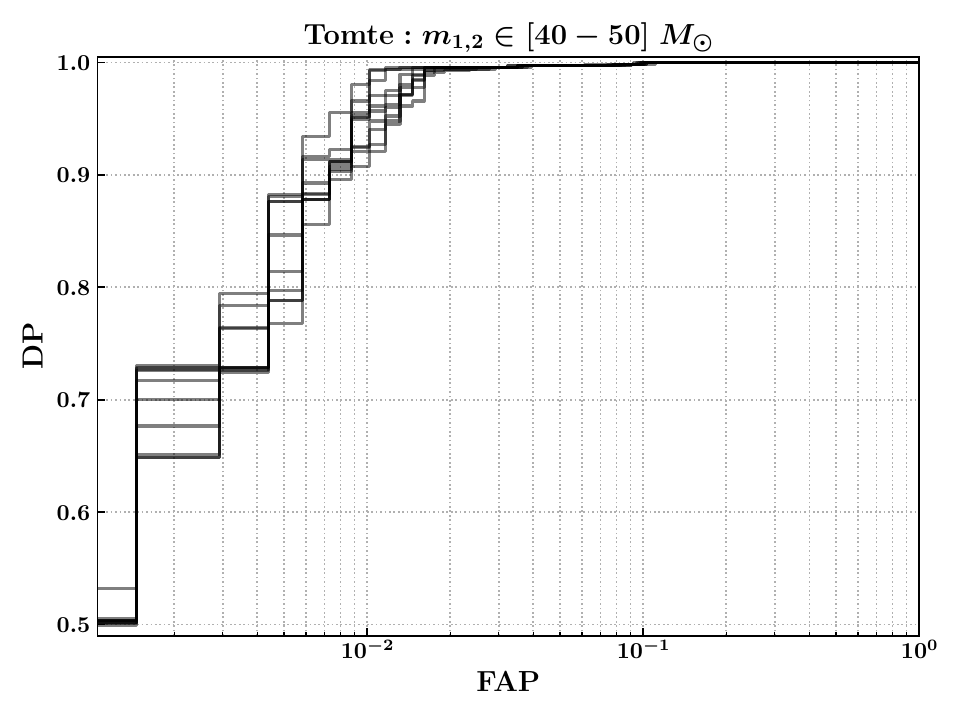}
    \caption{ROC curves demonstrating the robustness of the tomte $\chi^{2}$ discriminator. Each curve is obtained using singular vectors constructed from one of $10$ independent subsets of $100$ tomte glitches (drawn with replacement).}
    \label{fig:roc_robust}
\end{figure*}

We further demonstrate the robustness of our method with tomte glitches by drawing $10$ independent subsets (with replacement) to construct the corresponding singular vectors, which are then used to compute the tomte $\chi^{2}$ for the representative mass bin $[40\text{--}50]~M_{\odot}$.
Figure~\ref{fig:roc_robust} shows that the resulting ROC curves are in close agreement with one another, despite being constructed from different realizations of the singular-vector basis. Although we illustrate this robustness for a single representative case, similar consistency is expected for other glitch classes and across different mass bins.
It is also important to note that sine-Gaussian waveforms are effective models for characterizing glitches considered in this study.

\begin{figure}
    \centering
    \includegraphics[scale=0.54]{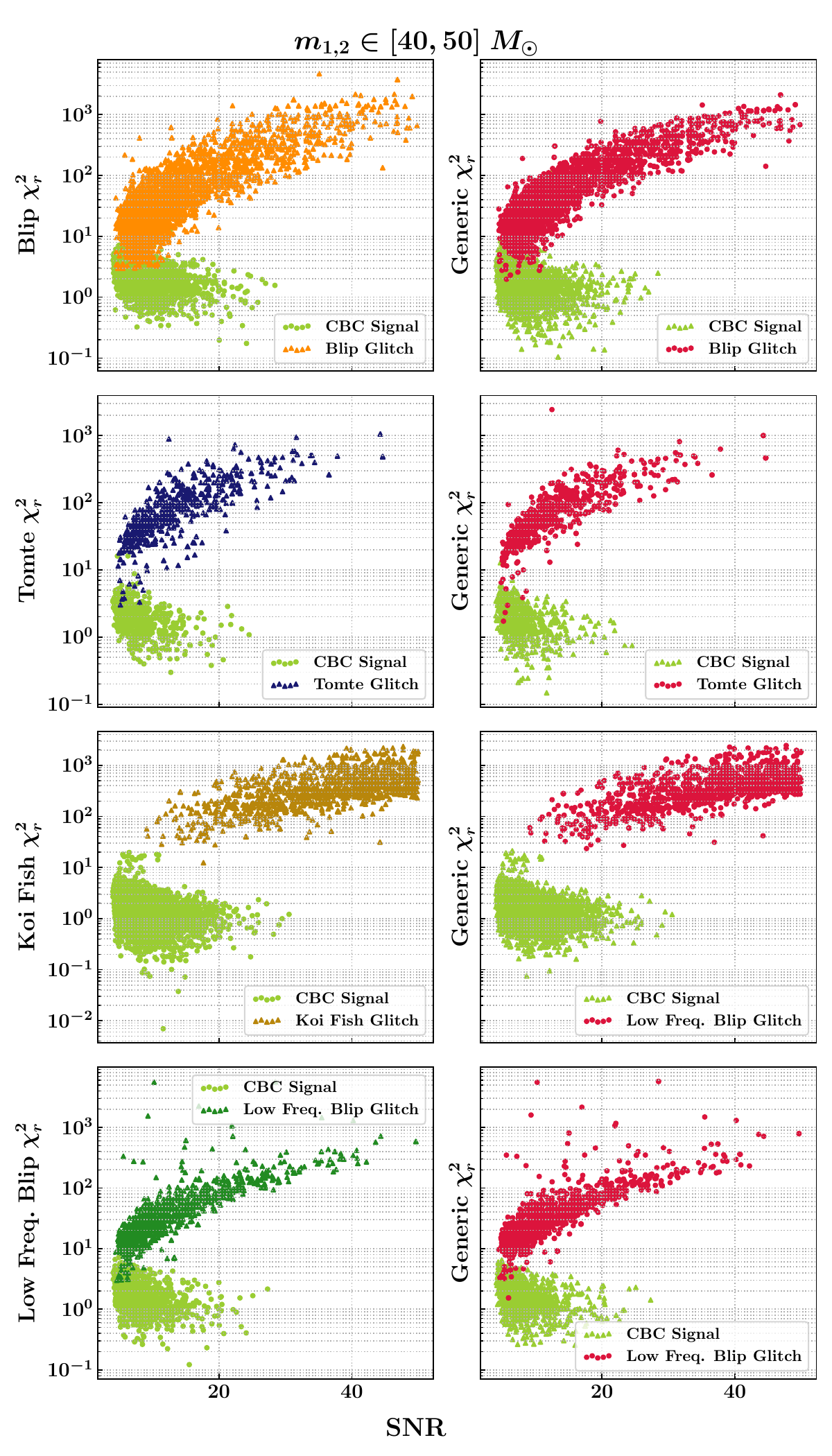}
    \caption{The left panel shows the glitch-specific $\chi^{2}$ {\it versus} SNR for both glitches and CBC signals, while the right panel presents the generic $\chi^{2}$ computed for the same glitches and simulated signals. Each row corresponds to a specific glitch type. The SNR of each data segment -- whether containing a CBC signal or a glitch -- is computed by projecting onto all CBC templates in the $40-50~M_{\odot}$ mass bin and taking the maximum value.}
    \label{fig:blip_chisq}
\end{figure}

The SNR time series, as shown in the top panel of figure~\ref{fig:cbc_trigger_time}, resulting from matched filtering of the CBC signal with the triggered CBC template, exhibits similarity with the SNR time series in the bottom panel of figure~\ref{fig:blip_trigger_time}, which results from the matched filtering of the singular vectors with the blip glitch. This similarity, manifesting as a sharp peak at the chirp signal or glitch time, indicates a significant overlap between the CBC template and singular vectors with the CBC signal and the blip glitch signal, respectively. 
However, when the CBC template is employed for matched filtering with the blip glitch, the peak of the SNR time series is broader compared to that obtained by matched filtering the singular vectors of the same blip glitch (see figure~\ref{fig:blip_trigger_time}). 
Similarly, the SNR time series peak from matched filtering the singular vectors with the CBC signal has a broader width than the peak of the SNR time series obtained by matched filtering of the CBC template with the CBC signal, as illustrated in figure~\ref{fig:cbc_trigger_time}.
This characteristic is also evident in the matched filter SNR values. The matched filter SNR between the CBC signal and the triggered CBC template is generally higher than the matched filter SNR between the CBC signal and the singular vectors. Similarly, the matched filter SNR between the blip and the singular vectors is usually greater than the matched filter SNR between the blip glitch and the triggered CBC template. Similar characteristics are also observed for tomte glitches.

The glitch-specific $\chi^{2}$ statistics proposed in this study prove effective due to the representation of the time-frequency morphology of the considered glitch classes with very few corresponding singular vectors.
In figure~\ref{fig:sv_H1}, we show the first $3$ singular vectors from $100$ glitches from each class observed in the LIGO-Hanford detector during the O3a run. 
Additionally, the matched filter SNR calculated between singular vectors and glitches from each class is significantly high (e.g., see figure~\ref{fig:blip_trigger_time} for a representative blip glitch ). So, singular vectors are good basis vectors for identifying the glitches from the corresponding class. 
However, the singular vectors are also triggered by the CBC signal. So, we have subtracted the triggered template from the singular vector to minimize the projection of the CBC signal onto the resultant vector. The effectiveness of this subtraction is evident in the low $\chi^{2}$ for the CBC signal, contrasting with the high $\chi^{2}$ for the glitches, as evident in figure~\ref{fig:blip_chisq}. This can be achieved due to the morphological dissimilarity between the CBC signal and the glitch.

\section{Conclusion} \label{conclusion}

In this work, we have successfully demonstrated how SVD-based $\chi^{2}$ effectively identify the glitches against
CBC signals and compared their performance to that of the power $\chi^{2}$ and sine-Gaussian $\chi^{2}$. One of the most important observations of this work is the universality of the underlying basis vectors of the glitches; i.e., just a few basis vectors are sufficient to capture most of the power in glitches and resolve them. It is also important to note that the efficacy of the optimized sine-Gaussian 
$\chi^{2}$ in identifying the glitches is quite similar to that of the glitch-specific and generic $\chi^{2}$. 
This also shows beyond doubt that the sine Gaussians are an excellent model of the glitches, hence, constitute an excellent choice as basis vectors for constructing statistics that effectively distinguish them from BBH signals.
This study provides confidence that the SVD method can be applied to model a variety of other glitches that affect the GW sensitivity to various astrophysical signals.
While the glitches considered in this work, were known to be well represented by sine-Gaussians, this may not be true for every glitch, but wherever SVD can be employed to accurately model them to a high degree of completeness, discriminators can be developed by following a similar procedure.
An important methodological consideration of the SVD-based approach is that the basis vectors are constructed using a set of well-characterized glitches from real detector data. In this work, we address this by constructing the singular vectors using glitches from the early part of the O3a run and applying them to glitches from the full O3a period. The consistent performance observed across all glitch classes indicates that the resulting singular vectors capture characteristic morphological features, rather than being tuned to a specific data segment. This behavior suggests that the method generalizes well within an observing run and supports its applicability to near–real-time analyses, where a limited set of previously identified glitches can be used to construct effective discriminators for subsequent data.

It is interesting to note that even in relatively lower-mass BBH searches, the glitch-specific $\chi^2$ does better than the power $\chi^2$. This has not been demonstrated before for the lower-mass bin studied here. It happens because, unlike the latter $\chi^2$, the former is indeed optimized for discrimination of BBHs against blips and tomtes. 

It must be noted that {\it prima facie} the computational cost of calculating the SVD-based $\chi^{2}$ is higher than that of the power $\chi^{2}$ and sine-Gaussian $\chi^{2}$. This is because we must always subtract the triggered BBH template from the singular vectors, followed by the Gram-Schmidt orthogonalization. However, the computational cost can be reduced by precomputing and storing the final resultant vectors for the entire CBC bank~\cite{Joshi:2020eds, Choudhary:2022nvs}. This and other investigations for improving the efficiency of computing the SVD-based $\chi^{2}$ may be pursued elsewhere.

\ack{}

The authors thank Yu-Chiung Lin for carefully reviewing the manuscript and providing several useful suggestions.
This work utilized the LDG cluster, Sarathi, at IUCAA for computational purposes.
T.G. acknowledges the support from JSPS Grant-in-Aid for Transformative Research Areas (A) No. 23H04893.
S.B. acknowledges support from the NSF under Grant PHY-2309352.

This research has made use of data or software obtained from the Gravitational Wave Open Science Center (gwosc.org), a service of the LIGO Scientific Collaboration, the Virgo Collaboration, and KAGRA. This material is based upon work supported by NSF's LIGO Laboratory which is a major facility fully funded by the National Science Foundation, as well as the Science and Technology Facilities Council (STFC) of the United Kingdom, the Max-Planck-Society (MPS), and the State of Niedersachsen/Germany for support of the construction of Advanced LIGO and construction and operation of the GEO600 detector. Additional support for Advanced LIGO was provided by the Australian Research Council. Virgo is funded, through the European Gravitational Observatory (EGO), by the French Centre National de Recherche Scientifique (CNRS), the Italian Istituto Nazionale di Fisica Nucleare (INFN) and the Dutch Nikhef, with contributions by institutions from Belgium, Germany, Greece, Hungary, Ireland, Japan, Monaco, Poland, Portugal, Spain. KAGRA is supported by the Ministry of Education, Culture, Sports, Science and Technology (MEXT), Japan Society for the Promotion of Science (JSPS) in Japan; the National Research Foundation (NRF) and Ministry of Science and ICT (MSIT) in Korea; Academia Sinica (AS) and National Science and Technology Council (NSTC) in Taiwan.

\section*{Data availability statement}
The data that support the findings
of this study are available upon reasonable request from the authors.

\bibliography{references}

\end{document}